\documentclass[12pt]{article}
\usepackage[latin1]{inputenc}
\usepackage[british]{babel}
\usepackage{cmap}
\usepackage{lmodern}

\usepackage{amssymb, amsmath, amsthm}
\usepackage[a4paper,top=25mm,bottom=25mm,left=25mm,right=25mm]{geometry}
\usepackage{etex}
\usepackage{ragged2e}

\usepackage{authblk} 
\usepackage{pifont}
\usepackage{graphicx}
\usepackage[usenames,dvipsnames,svgnames,table]{xcolor}
\usepackage[figuresright]{rotating}
\usepackage{xtab} 
\usepackage{longtable} 
\usepackage{multirow}
\usepackage{footnote}
\usepackage[stable]{footmisc}
\usepackage{chngpage} 
\usepackage{pdflscape} 
\usepackage{tocbibind} 

\usepackage{pgfplots}
\pgfplotsset{compat=1.14}
\usepackage{setspace}

\makesavenoteenv{tabular}
\usepackage{tabularx}
\usepackage{booktabs}
\usepackage{threeparttable}
\usepackage[referable]{threeparttablex} 
\newcolumntype{R}{>{\raggedleft\arraybackslash}X}
\newcolumntype{L}{>{\raggedright\arraybackslash}X}
\newcolumntype{C}{>{\centering\arraybackslash}X}
\newcolumntype{A}{>{\columncolor{gray!25}}C}
\newcolumntype{a}{>{\columncolor{gray!25}}c}

\newlength{\tablen}

\usepackage{dcolumn} 
\newcolumntype{.}{D{.}{.}{-1}}

\usepackage{tikz}
\usetikzlibrary{arrows}
\usepackage[semicolon]{natbib}
\usepackage[hyphens]{url}
\usepackage{hyperref} 
\hypersetup{
  colorlinks   = true,    
  urlcolor     = blue,    
  linkcolor    = blue,    
  citecolor    = red      
}
\usepackage{microtype}
\usepackage[justification=centering]{caption} 

\usepackage[labelformat=simple]{subcaption}

\DeclareCaptionLabelFormat{parenthesis}{(#2)}
\makeatletter
\renewcommand\p@subfigure{\arabic{figure}.}
\makeatother

\DeclareCaptionLabelFormat{parenthesis}{(#2)}
\makeatletter
\renewcommand\p@subtable{\arabic{table}.}
\makeatother

\usepackage{enumitem}

\setlist[itemize]{leftmargin=2.5\parindent}
\setlist[enumerate]{leftmargin=2.5\parindent}

\theoremstyle{plain}

\theoremstyle{definition}

\theoremstyle{remark}

\def\keywords{\vspace{.5em} 
{\noindent \textit{Keywords}:\,}}

\def\JEL{\vspace{.5em} 
{\noindent \textbf{\emph{JEL} classification number}:\,}}

\def\AMS{\vspace{.5em} 
{\noindent \textbf{\emph{MSC} class}:\,}}

\author{\href{https://sites.google.com/site/laszlocsato87}{L\'aszl\'o Csat\'o}\thanks{~E-mail: \emph{laszlo.csato@uni-corvinus.hu}} }
\affil{Institute for Computer Science and Control, Hungarian Academy of Sciences (MTA SZTAKI) \\
Laboratory on Engineering and Management Intelligence, Research Group of Operations Research and Decision Systems}
\affil{Corvinus University of Budapest (BCE) \\
Department of Operations Research and Actuarial Sciences}
\affil{Budapest, Hungary}
\title{UEFA Champions League entry has not satisfied strategyproofness in three seasons}
\date{\today}

\def\Dedication{
{\noindent
$\mathfrak{Alle}$ $\mathfrak{diese}$ $\mathfrak{Kenntnisse}$ $\mathfrak{lassen}$ $\mathfrak{sich}$ $\mathfrak{nicht}$ $\mathfrak{durch}$ $\mathfrak{den}$ $\mathfrak{Apparat}$ $\mathfrak{wissenschaftlicher}$ $\mathfrak{Formeln}$ $\mathfrak{und}$ \linebreak $\mathfrak{Maschinerien}$ $\mathfrak{erzwingen,}$ $\mathfrak{sondern}$ $\mathfrak{sie}$ $\mathfrak{erwerben}$ $\mathfrak{sich}$ $\mathfrak{nur,}$ $\mathfrak{wenn}$ $\mathfrak{in}$ $\mathfrak{der}$ $\mathfrak{Betrachtung}$ $\mathfrak{der}$ $\mathfrak{Dinge}$ $\mathfrak{und}$ $\mathfrak{im}$ $\mathfrak{Leben}$ $\mathfrak{ein}$ $\mathfrak{treffendes}$ $\mathfrak{Urteil,}$ $\mathfrak{wenn}$ $\mathfrak{ein}$ $\mathfrak{nach}$ $\mathfrak{dieser}$ $\mathfrak{Auffassung}$ $\mathfrak{hingerichtetes}$ $\mathfrak{Talent}$ $\mathfrak{t\ddot{a}tig}$ $\mathfrak{ist}$.}\footnote{~``\emph{These are things the knowledge of which cannot be forced out by an apparatus of scientific formula and machinery: they are only to be gained by the exercise of an accurate judgment in the observation of things and of men, aided by a special talent for the apprehension of both.}'' (Source: Carl von Clausewitz: \emph{On War}, translated by Colonel James John Graham, London, N. Tr\"ubner, 1873. \url{http://clausewitz.com/readings/OnWar1873/TOC.htm})}
\vspace{0.25cm}

\flushright
\noindent (Carl von Clausewitz: \emph{Vom Kriege})

\vspace{1cm} 
\justify }

\begin{document}

\maketitle

\Dedication

\begin{abstract}
\noindent
The paper investigates the qualification for the UEFA Champions League, the most prestigious club competition in European football with respect to the theoretical property of strategy-proofness.
We find that in three seasons (2015-16, 2016-17, 2017-18), the UEFA Europa League titleholder might have been better off by losing its match against the Champions League titleholder in their domestic championship. A straightforward solution is suggested in order to avoid the occurrence of this paradox. The use of an incentive compatible rule would have a real effect on the qualification in these three seasons of the UEFA Champions League.

\JEL{C44, D71, Z20}

\AMS{62F07, 91B14}

\keywords{OR in sports; tournament ranking; UEFA Champions League; strategy-proofness; manipulation}
\end{abstract}

\section{Introduction} \label{Sec1}

In an appropriately designed tournament, players are interested in eliciting costly effort to win as many games as possible. However, sometimes a team might be punished for showing a better performance \citep{KendallLenten2017}.

This note, similarly to some recent works of the field \citep{DagaevSonin2018, Csato2018h, Csato2018b, Csato2019f}, focuses on the particular case when a team is guaranteed to be better off by losing a match, that is, the probabilistic aspect of manipulation \citep{Pauly2014, Vong2017} is neglected. Specifically, we will show that UEFA Champions League (CL) entry has been incentive incompatible in the three seasons between the years 2015 and 2018.

The problem is caused by the rule describing the qualification of the UEFA Europa League (EL) titleholder (from the previous season) for the CL: ``\emph{The UEFA Europa League titleholder is guaranteed a place in the competition as a minimum in the play-offs. It will have priority filling a vacancy created in the group stage or in the play-offs by the UEFA Champions League titleholder}'' (see Article~3.04 of \citet{UEFA2015a}, \citet{UEFA2016d}, and \citet{UEFA2017b} for the 2015-16, 2016-17, and 2017-18 seasons, respectively).
Consequently, since a place in the group stage is preferred to a place in the play-offs, the EL titleholder is interested in creating a vacancy in the group stage, which can be achieved in the domestic championship if EL titleholder is from the same association as the CL titleholder.

Based on a theoretical finding of \citet{DagaevSonin2018}, we suggest a slight modification in filling the potential vacancy to guarantee strategy-proofness.
The use of an incentive compatible rule would have a real effect on the qualification in these three seasons of the Champions League.

Nevertheless, it should be mentioned that the situation to be described here has many conditions and they have never been fulfilled. Even in this case, probably no team would have been an incentive to lose intentionally a match because winning would have a higher expected value. Furthermore, the controversial rule is not applied currently by the UEFA.
On the other hand, even an improbable scenario may cause problems in practice (for instance, \citet{Csato2019h} shows that the theoretical issue outlined in \citet{DagaevSonin2018} has arisen in the 2011-12 season of the Dutch national soccer championship), and punishing a team for its better performance seems to be a severe violation of fairness.

The rest of the article is organized as follows.
Section~\ref{Sec2} presents a real-world illustration of the problem.
Its background is discussed in Section~\ref{Sec3}, while Section~\ref{Sec4} concludes.

\section{A hypothetical example} \label{Sec2}

The Premier League -- the top English professional league for association football clubs, played as a home-away round-robin tournament -- ranks the teams lexicographically with the number of points being the first criterion. A win is awarded by three points, and a draw is awarded by one point.

On the basis of the results in the 2016-17 season, England has had four places in the 2017-18 UEFA Champions League allocated as follows \citep[Annex~A]{UEFA2017b}:
\begin{itemize}
\item
the winner, the runner-up, and the third-placed club qualify for the CL group stage; and
\item
the fourth-placed club qualifies for the CL play-off.
\end{itemize}

Hence, if the CL titleholder is among the best three clubs, then the EL titleholder qualifies for the group stage of the CL, as described in Section~\ref{Sec1}. Otherwise, it should play a play-off, and it advances to the CL group stage only by winning this particular play-off over to legs against a team from another UEFA association.

\begin{table}[ht!]
\begin{threeparttable}
\centering
\caption{Final ranking of the 2016-17 Premier League season}
\label{Table1}
\rowcolors{1}{gray!20}{}
    \begin{tabularx}{\linewidth}{Cl CCC CCC >{\bfseries}C} \toprule \hiderowcolors
    Pos   & Team  & W     & D     & L     & GF    & GA    & GD    & Pts \\ \hline \showrowcolors
    1     & \textbf{Chelsea} & 30    & 3     & 5     & 85    & 33    & 52    & 93 \\
    2     & \textbf{Tottenham Hotspur} & 26    & 8     & 4     & 86    & 26    & 60    & 86 \\
    3     & \textbf{Manchester City} & 23    & 9     & 6     & 80    & 39    & 41    & 78 \\
    4     & \textit{Liverpool} & 22    & 10    & 6     & 78    & 42    & 36    & 76 \\
    5     & Arsenal & 23    & 6     & 9     & 77    & 44    & 33    & 75 \\
    6     & \textbf{Manchester United} & 18    & 15    & 5     & 54    & 29    & 25    & 69 \\ \hline    
    \end{tabularx}
    
    \begin{tablenotes}
\item
\footnotesize{Pos = Position; W = Won; D = Drawn; L = Lost; GF = Goals for; GA = Goals against; GD = Goal difference; Pts = Points. \\
There are three types of teams. Some of them go to the Champions League group stage (in bold), some of them go to the Champions League play-offs (italicized) and some of them do not qualify for the Champions League (roman).}   
    \end{tablenotes}
\end{threeparttable}
\end{table}

Table~\ref{Table1} shows the final league table for the top six teams at the end of the 2016-17 season. 
Consider the following scenario. Suppose that Manchester City have won the 2016-17 UEFA Champions League, and Manchester United have won the 2016-17 UEFA Europa League.\footnote{~Actually, Manchester United have won the EL, but Manchester City have been eliminated in the round of 16 of the CL.}
According to the entry rules, the CL titleholder Manchester City qualify for the group stage of the CL through its domestic championship, which creates a vacancy in the group stage, filled by the EL titleholder Manchester United.

Manchester United have lost by 1-2 at home against Manchester City on 10 September 2016.
What would have happened if Manchester United would have defeated Manchester City in this match? Then Manchester City would stand with 75 points, so Liverpool would be the third at the end of the 2016-17 season and would qualify for the CL group stage. Furthermore, Manchester City would qualify for the CL group stage as the titleholder, however, Manchester United would qualify only for the CL play-offs, and should defeat an opponent in order to qualify for the group stage \citep[Article~3.03]{UEFA2017b}.

To conclude, Manchester United would be strictly worse off (at least, with respect to the CL qualification) if it would have defeated Manchester City.

It may seem at a first sight that this paradox is almost irrelevant since its occurrence assumes full knowledge of the future. Nevertheless, we think it means a serious violation of fairness because the misaligned UEFA rule punishes Manchester United for its better results.

\section{Discussion} \label{Sec3}

The incentive incompatible allocation rule has been used in three CL seasons (2015-16, 2016-17, 2017-18).
Then the scenario presented in Section~\ref{Sec2} would have occurred when:
\begin{itemize}
\item
the two (CL and EL) titleholders are from the same national association; and
\item
at least one team is directly qualified for the CL group stage from this national association.
\end{itemize}

Hence, according to the access list of the 2017-18 CL season \citep[Annex~A]{UEFA2017b}, the problem concerns the twelve strongest associations: Spain, Germany, England, Italy, Portugal, France, Russia, Ukraine, Belgium, Netherlands, Turkey, and Switzerland.\footnote{~The set of the top twelve strongest associations was the same in the 2016-17 season \citep[Annex~A]{UEFA2016d}, but Greece was among them instead of Switzerland in the 2015-16 season \citep[Annex~A]{UEFA2015a}.}

In these years the CL titleholders were Barcelona, Real Madrid (both from Spain), and Real Madrid, respectively, while the EL titleholders were Sevilla (from Spain), Sevilla, and Manchester United, respectively. Despite the CL and EL titleholders were from the same UEFA association in two seasons, there was no danger of manipulation since the CL titleholders were safely among the top teams that directly qualify for the CL group stage.

A situation close to the one described in Section~\ref{Sec2} has been modelled by \citet{DagaevSonin2018} in general. \citet[Proposition~3]{DagaevSonin2018} can be replicated when the domestic championship is a round-robin tournament played in two rounds on a home-away basis, while the UEFA Champions League and the UEFA Europa League are considered as knock-out tournaments. This result practically says that the qualification is strategy-proof if and only if all vacant slots are allocated on the basis of the round-robin tournament, i.e., the domestic championship.

In our real-world problem, the teams compete for slots in two types of tournaments, the group stage and the play-offs of the CL, where the former is more valuable. It means that, similarly to \citet[Section titled ``Extensions and Discussion'']{DagaevSonin2018}, a general formal analysis would be cumbersome as the number of types of vacancies increases dramatically (we have even two knock-out tournaments instead of one).
Thus, instead of analysing all possible allocation rules, we consider only the actual UEFA regulation, which gives a priority for the EL titleholder to fill any vacancy in the CL group stage created by the CL titleholder.

Then \citet[Proposition~3]{DagaevSonin2018} implies that the vacant slot should be allocated on the basis of the round-robin domestic championship in order to avoid a possible punishment of the EL titleholder. We think it is the obvious mechanism that guarantees incentive compatibility (at least with respect to the qualification of the EL titleholder).

This strategy-proof version of the allocation rule -- which fills the vacancy from the domestic championship -- would make a real difference in these three CL seasons:
\begin{itemize}
\item
2015-16: Valencia (the fourth team in Spain) would have qualified for the CL group stage instead of Sevilla (EL titleholder), which would have played the CL play-off against Monaco (from France); 
\item
2016-17: Villarreal (the fourth team in Spain) would have qualified for the CL group stage instead of Sevilla (EL titleholder), which would have played the CL play-off against Monaco; 
\item
2016-17: Liverpool (the fourth team in England) would have qualified for the CL group stage instead of Manchester United (EL titleholder), which would have played the CL play-off against 1899 Hoffenheim (from Germany).
\end{itemize}

Consequently, Villarreal would not have suffered a significant financial loss due to its elimination from the CL group stage in the 2016-17 season.

Despite a situation susceptible to manipulation has not materialized, the potential betting markets and match-fixing implications could be quite severe. Furthermore, it seems unfair to use an allocation rule that may punish a team for its better performance. Perhaps UEFA administrators have recognized this danger: in the framework of a substantial reform of the CL qualification, they have decided to guarantee a slot for the EL titleholder in the CL group stage from the 2018-19 season \citep{UEFA2018b}.

While it clearly eliminates the problem of bad incentives, the current mechanism forces another team to play a play-off instead of the EL titleholder in order to qualify for the CL group stage. Hence its superiority or inferiority compared to our suggestion depends on a policy choice.

\section{Conclusions} \label{Sec4}

Regulations governing major sports are usually thought to be relatively stable. In fact, some rule books are under constant development, for example, the UEFA Champions League entry is revised in every three years. These changes sometimes have unforeseen consequences as illustrated above. 

We think that even the marginal probability of a sports ranking rule working imperfectly is a sufficient reason for scientific researchers to write notes and papers such as the current one in order to report these issues and suggest ways to circumvent them.

\section*{Acknowledgements}
\addcontentsline{toc}{section}{Acknowledgements}
\noindent
We are grateful to \emph{M\'at\'e Stift} for raising the issue of the qualification of titleholders at the end of a presentation. \\
We are indebted to the \href{https://en.wikipedia.org/wiki/Wikipedia_community}{Wikipedia community} for contributing to our research by collecting and structuring information used in the paper. \\
The research was supported by OTKA grant K 111797 and by the MTA Premium Post Doctorate Research Program. 

\bibliographystyle{apalike}
\bibliography{All_references}

\end{document}